\documentclass[12]{svmult}

\usepackage{makeidx}         
\usepackage[pdftex]{graphicx}        
\usepackage{multicol}        
\usepackage[square,numbers]{natbib}

\usepackage[pdftex]{color}

\newcommand{\av}[1]{\langle {#1} \rangle}

\makeindex             

\begin{document}

\title*{Generalized voter-like models on heterogeneous networks}
\author{Paolo Moretti\inst{1} \and Andrea Baronchelli\inst{2} \and
  Michele Starnini\inst{3} \and Romualdo Pastor-Satorras\inst{3}}

\institute{Departamento de Electromagnetismo y F\'isica de la Materia and\\
Instituto ``Carlos I'' de F\'isica Te\'orica y Computacional, \\
Universidad de Granada,
Facultad de Ciencias,
Fuentenueva s/n, E-18071 Granada,
Spain
  \and Department of Physics, College of Computer and
  Information Sciences, Bouv\'e College of Health Sciences,
  Northeastern University, Boston MA02120, USA
  \and Departament de F\'\i sica i Enginyeria Nuclear,
Universitat Polit\`ecnica de Catalunya, Campus Nord B4, 08034
Barcelona, Spain}
%
%

\authorrunning{P. Moretti, A. Baronchelli, M. Starnini and R. Pastor-Satorras}
\maketitle

\section{Introduction}
\label{sec:introduction}

The study of dynamical ordering phenomena and consensus formation in
initially disordered populations is a central problem in the
statistical physics approach to social and natural sciences, whether
the broad idea of consensus is referred to opinions, voting
intentions, language conventions, social habits or inherited genetic
information. Indeed, people tend to align their opinions
\cite{Castellano09}, segregated populations gradually lose their
genetic diversity \cite{Crowbook}, and different social groups
spontaneously develop their own seemingly arbitrary traits of dress or
jargon \cite{barth1969ega}. Understanding how global order can emerge
in these situations, self-organized by purely local interactions,
represents an important theoretical and practical problem.  Our
ability to grasp these issues has been mainly driven by the analysis
of simple statistical models, which capture the essential ingredients
of a copying/invasion local dynamics. The simplest of those
copying/invasion processes are the voter model \cite{Clifford73} and
the Moran process \cite{moran1958random}, which focus principally on
social \cite{Castellano09} and evolutionary \cite{nowak2006ed}
dynamics, respectively.

These two models describe a population as a set of agents, each one
carrying a state (opinion, trait, genome) represented by a binary
variable $\sigma=\pm1$.  At each time step, an ordered pair of
adjacent agents $(i,j)$ is selected at random. In the voter model, as
a paradigm of copying processes, the system is updated as $\sigma_i :=
\sigma_j$, the first agent copying the state of its neighbor. The
voter model can thus be conceived a very simplistic model of opinion
formation in society in which individuals select their beliefs by the
(admittedly not very realistic) procedure of just imitating one of
their neighbors.  On the other hand, in the Moran process it is the
neighboring agent the one who copies the state of the first agent,
$\sigma_j := \sigma_i$, or, from another perspective, the state of the
first one ``invades'' the neighbor in contact. The Moran process
represents thus a simple approximation to the evolutionary dynamics of
a haploid population, constrained to have a fixed size
\cite{Crowbook}.

In finite, initially disordered systems, and in the absence of bulk
noise (i.e. agents spontaneously changing their state
\cite{Castellano09}), stochastic copying/invasion dynamics lead to a
uniform state with all individuals sharing the same value $\sigma$,
the so-called consensus. Being the two final states symmetrical in
principle, the dynamical evolution towards either state will only depend on
the initial configuration. In order to characterize how the consensus
is reached, two quantities are usually consider. The exit probability
$E(x)$ is defined as the probability that the final state corresponds
to all agents in the state $+1$ when starting from a homogeneous
initial condition with a fraction $x$ of agents in state
$+1$. Accordingly, the consensus time $T_N(x)$ is defined as the
average time required to reach consensus, independently of its value,
in a system of $N$ agents.

Voter-like processes were originally considered in regular topologies.
In this case, the voter model remarkably is one of the few stochastic
non-equilibrium models that can be exactly solved in any number of
dimensions \cite{liggett99:_stoch_inter,PhysRevE.53.R3009}. As it
turns out, its symmetries imply the ensemble conservation of
magnetization $m = \sum_i \sigma_i /N$, which in turn implies that the
exit probability takes the linear form $E(x) = x$. On the other hand,
the consensus time can be seen to scale with the number $N$ of agents
as $T_N \sim N^2$ in $d=1$, $T_N \sim N \ln(N)$ in $d=2$ and $T_N \sim
N$ for $d>2$ \cite{KineticViewRedner}.

The results of the voter and other copying/invasion processes in the
context of social and evolutionary dynamics acquire a larger relevance
when they are considered in systems endowed with realistic,
non-trivial topologies.  In fact, the strong heterogeneity of social
and ecological substrates is better encoded in terms of a complex
network \cite{Newman2010}, rather than by a homogeneous
$d-$dimensional lattice. Therefore, a large theoretical effort has
been recently devoted to uncover the effects of a complex topology on
the behavior of the voter model, as well as general dynamical
processes \cite{dorogovtsev07:_critic_phenom,barratbook}. At the most
basic level, while voter and Moran models are equivalent at the
mean-field level for regular topologies, if the connection pattern is
given by a complex network they behave differently, since the order in
which the interacting agents $(i,j)$ are selected becomes
relevant~\cite{Suchecki05,castellano05:_effec}. Additionally, the
heterogeneity in the connection pattern, as measured by the degree
distribution $P(k)$, defined as the probability that a randomly chosen
agent is connected to $k$ other agents (has degree $k$) plays a
relevant role in the scaling of the consensus time, especially for
scale-free networks \cite{Barabasi:1999} with a degree distribution
scaling as a power-law, $P(k)\sim k^{-\gamma}$
\cite{Suchecki05,Castellano05,castellano05:_effec}. A theoretical
understanding of the behavior of the voter model on heterogeneous
networks was finally put forward by Redner and coworkers
\cite{PhysRevLett.94.178701,Antal06,Sood08}, who showed that for the
voter model the consensus time scale with the system size as $T_N \sim
N \av{k}^2 / \av{k^2}$, i.e. inversely proportional to the second
moment of the degree distribution $\av{k^2} = \sum_k k^2 P(k)$. In
scale-free networks with degree exponent $\gamma <3$, the second
moment $\av{k^2}$ diverges and becomes size dependent, implying thus a
sublinear growth of the consensus time, as previously observed in
numerical simulations \cite{Suchecki05,Castellano05}, and contrarily
to the homogeneous mean-field expectation.

When discussing the role of heterogeneity in social or evolutionary
contexts, a relevant question is whether the complexity of the
substrate alone is able to encode the heterogeneity of a realistic
dynamical process of social or environmental relevance. The common
objection to extremely simple models is that in reality individuals
behave and relate to their peers in different ways, i.e. they are
\textit{heterogeneous} both in the way in which they interact with
others and in the way in which they react to these interaction. For
example, in a social context, it could be the case that some agents
are more reluctant to change their opinion (\textit{zealots}), while
some agents can assign different importance to the opinion of their
different neighbors (i.e. close friends can be more trusted than
casual acquaintances). Lately, different variations of voter-like
models have been put forward, in an effort to take into account the
intrinsic variability of agents and their individual propensity to
interact with peers
\cite{Antal06,schneider-mizell09,Yang09,Lin10,2010arXiv1011.2395B}.

In this chapter, we describe a generalization of the voter model on
complex networks that encompasses different sources of degree-related
heterogeneity and that is amenable to direct analytical solution by
applying the standard methods of heterogeneous mean-field theory
\cite{dorogovtsev07:_critic_phenom,barratbook}. Our formalism allows
for a compact description of previously proposed heterogeneous
voter-like models, and represents a basic framework within which we
can rationalize the effects of heterogeneity in voter-like models, as
well as implement novel sources of heterogeneity, not previously
considered in the literature.

\section{A generalized heterogeneous voter-like model on networks}
\label{sec:gener-voter-model}

We consider a generalized heterogeneous voter-like model, given as a
stochastic process on networks, and defined by the following rules
\cite{morettiheterovoter}:
\begin{itemize}
\item Each vertex $i$ has associated a given \textit{fitness} $f_i$.
\item A source vertex $i$ is selected at random, with a probability
  $f_i / \sum_j f_j$, i.e., proportional to its fitness $f_i$.
\item A nearest neighbor $j$ of $i$ is then selected at random.
\item With probability $Q(i,j)$, $i$ copies the state of vertex $j$
  with. Otherwise, nothing happens.
\end{itemize}

The fitness function $f_i$ affects the probability that the given node
$i$ is chosen to initiate the opinion-update process at a given time
$t$ \cite{Antal06}.  In the case of copying dynamics, it measures the
propensity of a given node to change its state. Once the
individual $i$ is chosen and a neighbor $j$ is selected, the probability
$Q(i,j)$ measures how probable the actual update is, introducing a
weight in the adjacency relation between the two individuals
\cite{2010arXiv1011.2395B}.  We can easily check that the standard
voter model is recovered by setting $f_i=Q(i,j)=1$, while the Moran
process corresponds to $f_i=1$ and $Q(i,j)=k_i/k_j$. Other variations of
copying/invasion dynamics can be recovered by the appropriate
selection of the $f_i$ and $Q(i,j)$ functions.

Recently, a generalized formalism for the class of copying/invasion
voter-like models on networks has been proposed
\cite{baxter:258701,1751-8121-43-38-385003} in which the process is
identified by the copying rate $C_{ij}$, encoding the full structure
of the contact network and the stochastic update rules, and that in
our case takes the form
\begin{equation}
  \label{eq:13}
  C_{ij} = \frac{f_i}{\sum_p f_p} \frac{a_{ij}}{k_i} Q(i,j),
\end{equation}
where $a_{ij}$ is the adjacency matrix of the network, taking value
$1$ if vertices $i$ and $j$ are connected by an edge, and zero
otherwise. Within this formalism, exact results can be obtained, but
at the expense of computing the spectral properties of the copying
rate matrix $C_{ij}$, a highly non-trivial task unless the matrix
$C_{ij}$ has a relatively simple form.

Here we follow a different path, applying to our model the technique
of heterogeneous mean-field theory, which leads to simple estimates
for central properties such as the exit probability and the consensus
time in a rather economical way. While this technique is known to be
not exact in several instances, it still nevertheless able to account
with a reasonable accuracy for the results of direct numerical
simulations of the model.

\section{Heterogeneous mean-field analysis}
\label{sec:heter-mean-field}

The analytical treatment of voter-like models on complex networks is
made possible by the heterogeneous mean-field (HMF) approach, which
has traditionally provided a powerful analysis tool for dynamical
processes on heterogeneous substrates
\cite{barratbook,dorogovtsev07:_critic_phenom}. Two main assumptions
are made: (\emph{i}) Vertices are grouped into degree classes, that
is, all vertices in the same class share the same degree and the same
dynamical properties; (\emph{ii}) The real (\textit{quenched}) network
is coarse-grained into an \textit{annealed} one
\cite{dorogovtsev07:_critic_phenom}, which disregards the specific
connection pattern and postulates that the class of degree $k$ is
connected to the class of degree $k'$ with conditional probability
$P(k'|k)$ \cite{marian1}. In general, the HMF approach allows for
simple analytic solutions. In the case of voter-like models it has
proved remarkably powerful in estimating the quantities of interest,
showing reasonable agreement with numerical simulations in real
quenched networks
\cite{PhysRevLett.94.178701,Antal06,Sood08,schneider-mizell09,2010arXiv1011.2395B}.

Following the standard HMF procedure, we work with the degree-class
average of the fitness function and microscopic copying rate.
Averages are taken over the set of vertices with a given fixed degree,
i.e.
\begin{eqnarray}
  f_i  &\to& \frac{1}{N P(k)} \sum_{i \in k} f_i \equiv f_k,\\
  Q(i,j) &\to& \frac{1}{N P(k)}\frac{1}{N P(k')} \sum_{i \in k} \sum_{i \in k'}  Q(i,j) \equiv Q(k,k'),
\end{eqnarray}
where $i \in k$ denotes a sum over the degree class $k$ and $P(k)$ is
the network's degree distribution. Thus, $f_k$ represents the fitness
of individuals of degree $k$, assumed to be the same, depending only
on degree, for all of them, while $Q(k,k')$ represents the probability
that a vertex of degree $k$ copies the state of a vertex of degree
$k'$. Analogously, the contact pattern is transformed according to the
method described in Ref.~\cite{dynam_in_weigh_networ}, obtaining
\begin{equation}
  \frac{a_{ij}}{k_i} \to  \frac{[N P(k)]^{-1} \sum_{i \in k} \sum_{j \in k'}
    a_{ij}}{[N P(k)]^{-1}  \sum_{i \in k} \sum_{r} a_{ir}} \equiv P(k'|k).
\end{equation}
Disregarding the {\it microscopic} details of the actual contact pattern, 
our generalized voter model is thus defined in terms of the {\it mesoscopic} 
copying rate
\begin{equation}
  C(k, k')\equiv\frac{f(k)}{\langle f(k) \rangle} P(k'|k) Q(k,k').
\end{equation}

Here and in the following me adopt the convention $\langle \cdot
\rangle = \sum _k P(k) (\cdot ) $ In order to provide a quantitative
measure of the ordering process, we shall consider the time evolution
of the fraction of vertices of degree $k$ in the state $+1$, $x_k$.
Transition rates for $x_k$ will be given by the probability $\Pi(k;
\sigma)$ that a spin in state $\sigma$ at a vertex of degree $k$ flips
its value to $-\sigma$
\cite{PhysRevLett.94.178701,Sood08,2010arXiv1011.2395B} in a
microscopic time step.  It is easy to show that, from the definition
of the generalized voter model, such probabilities can be written as
\begin{eqnarray}
  \label{eq:15}
  \Pi(k;+1)&=&x_k P(k)\sum_{k'}(1-x_{k'})C(k,k')\\
  \Pi(k;-1)&=&(1-x_k)P(k)\sum_{k'}x_{k'}C(k,k'),
\end{eqnarray}
thus leading to the rate equation \cite{2010arXiv1011.2395B}
\begin{eqnarray}
  \dot{x}_k &=&\frac{\Pi(k; -1)-\Pi(k; +1)}{P(k)} 
	\equiv \sum_{k'}C (k,k') (x_{k'}-x_k).\label{eq:first}
\end{eqnarray}

Given the very broad definition of the interaction rate $Q(k,k')$, a
solution to the problem at hand cannot be easily provided in closed
form, unless further assumptions are made. We well thus make the
following additional assumptions:
\begin{itemize}
\item [\ ](\emph{i}) Dynamics proceed on uncorrelated networks,
  i.e. \cite{mendesbook}
  \begin{equation}
    P(k'|k)=\frac{k'P(k')}{\langle k \rangle};
  \end{equation}
\item [\ ](\emph{ii}) The interaction rate can be factorized as
\begin{equation}\label{eq:voterQ}
  Q(k,k')=a(k)b(k')s(k,k'),
\end{equation}
where $s(k,k')$ is any symmetric function of $k$ and $k'$.
\end{itemize}
This simplified form encompasses a broad range of voter-like dynamical
processes, including most previously proposed models and a variety of
novel applications, with the remarkable advantage of being 
promptly solvable, as it will become clear in the rest of this Chapter.

In order to provide a general solution in the most compact notation 
possible, we rewrite the rate equation as
\begin{equation}\label{eq:generalized}
\dot{x}_k=\sum_{k'}P(k')\Gamma (k,k') (x_{k'}-x_k),
\end{equation}
where we have defined $\Gamma(k,k')= u(k) v(k') s(k,k')$ and
\begin{equation}
  u(k)=\frac{a(k)f(k)}{\langle f(k)\rangle}, \qquad
  v(k')= \frac{b(k')k'}{\langle k \rangle}. \label{eq:1}
\end{equation}

We analyze the behavior of the linear process at hand in 
the canonical way, by first determining
the inherent conservation laws
\cite{PhysRevLett.94.178701,Antal06,Sood08}.
We define a generic integral of motion
$\omega[x_k(t)]$ for Eq.~(\ref{eq:generalized}),
 such that $d\omega/dt=0$. By definition of total time derivative
\begin{equation}
  \frac{d\omega}{dt}=\nabla_{\bf x}\omega \cdot {\bf \dot{x}}=\sum_k
  \frac{\partial \omega}{\partial x_k} \dot{x}_k=0. 
\end{equation}
In analogy with previous results
\cite{PhysRevLett.94.178701,Antal06,Sood08}, we look for conserved
quantities that are linear in $x_k$ imposing $\partial \omega/\partial
x_k=z_k$ independent of $x_k$, so that conserved quantities will be
given by
\begin{equation}
\omega={\bf z} \cdot {\bf x} =\sum_k z_k x_k,
\end{equation}
where $z_k$ is any solution of $\sum_k z_k \dot{x}_k=0$ and
$\dot{x}_k$ is given by Eq. (\ref{eq:generalized}).  
It is easy to prove that  $z_k\propto
P(k) v(k)/u(k)$ always satisfies the above condition, so that a
conserved quantity is found up to multiplicative factors and additive
constants.  Upon imposing $\sum_k z_k =1$ as one of the possible 
normalization conditions, the conserved quantity becomes
\begin{equation}
  \omega={\bf z} \cdot {\bf x} =\frac{\langle v(k)/
    u(k)\,x_k\rangle}{\langle v(k)/u(k)\rangle}.
\end{equation}
In analogy with the simplest definition of the voter model
~\cite{PhysRevLett.94.178701} the
conserved quantity bears all the information required to calculate 
the exit probability $E$, which we previously introduced as
 the probability that the final
state corresponds to all spins in the state $+1$.  In the final state
with all $+1$ spins we have $\omega=1$, while $\omega=0$ is the other
possible final state (all $-1$ spins).  Conservation of $\omega$
implies then $\omega = E \cdot 1+ [1-E] \cdot 0$, hence
\begin{equation}
  E = \omega = \frac{\langle v(k)/
    u(k)\,x_k\rangle}{\langle v(k)/u(k)\rangle}.
  \label{eq:3}
\end{equation}
Starting from a homogeneous initial condition, with a given density
$x$ of randomly chosen vertices in the state $+1$, we obtain, since
$\omega = x$,
\begin{equation}
  E_h(x) = x,
\end{equation}
completely independent of the defining functions $a$, $b$, and $s$, and
taking the same form as for the standard voter model
\cite{Castellano09}. On the other hand, if the initial condition corresponds 
to a single seed, that is an individual $+1$ spin in a vertex of degree $k$, 
\begin{equation}
  \omega= E_1(k) = \frac{v(k)/ u(k)}{N \langle v(k)/u(k)\rangle}, \quad 
\end{equation}
which does not depend on the functional form of the symmetric
interaction term $s(k,k')$.

Eq. (\ref{eq:generalized}) predicts that the set of variables of $x_k$ 
rapidly converge to a steady state. It is easy to see that any choice
of $x_k$ that is 
constant in $k$ is a solution to the steady state condition
$\dot{x}_k=0$. This solution is unique and does not
depend on initial conditions if the square matrix $P(k')\Gamma(k,k')$
is irreducible and primitive (it certainly is when working with
positive rates, which we will do in the following) \cite{gantmacher}.
If we call the steady state $x^\infty$, then it is easy to prove that
\begin{equation}
  \omega= \sum_{k'} z_{k'} x_{k'}=x^\infty,
\end{equation}
that is the steady state value for $x_k$
equals the conserved quantity. Such result is well-known in 
simpler formulations of the voter model and becomes crucial
in the computation of the consensus time, even in our
generalized case.

As we noted above, the convergence to the steady state
distribution occurs on very short time scales.
As soon as the steady
state is reached, stochastic fluctuations 
become relevant and the systems begins to fluctuate diffusively around this
value, until consensus is reached in one of the two symmetric states. 
Such fluctuations are integral to
finite systems and occur at long time scales, making this time-scale 
separation possible in large enough systems.  
In the light of such considerations, the average consensus time
$T_N({\bf x})$ for a system in a generic steady state ${\bf x}$ can be
derived extending the well known recursive method to our general case
\cite{Sood08}.  At a given time $t$, $T_N({\bf x})$ must equal the
average consensus time at time $t+\Delta t$ plus the elapsed time
$\Delta t=1/N$ that is, in our notation,
\begin{equation}
  T_N({\bf x})=\bar\Pi \,T_N({\bf x})+\sum_{k,s}\Pi(k;s)T_N({\bf x}+{\bf \Delta x}^{(k)})
  +\Delta t,
\end{equation}
where $\bar\Pi=1-\sum_{k,s}\Pi(k;s)$ is the probability that no state
change occurs, while the sum is the weighted average over possible
state-updates ${\bf x}\to {\bf x}+{\bf \Delta x}^{(k)}$. The variation
${\bf \Delta x}^{(k)}$ is a vector whose all components are zero
except for the $k$-th, which equals the update-unit
$\Delta_k=[NP(k)]^{-1}$. Expanding to second order in $\Delta_{k}$,
taking $x_k=\omega$ as the initial state and changing variables such
that $\partial/\partial x_k=z_k\partial/\partial\omega$ we obtain the
backward Kolmogorov equation
\begin{equation}
-1=\frac{ {\bf z}^T\Gamma{\bf z} }{N} \omega(1-\omega)\frac{\partial^2T_N}{
\partial \omega^2}
\end{equation}
leading to
\begin{equation}\label{eq:TN}
  T_N=-N_{\mathrm{eff}} [\omega \ln \omega + (1-\omega) \ln (1-\omega)]
\end{equation}
where we have defined the effective system size 
$N_{\mathrm{eff}} = N / \sum_{k,k'}z_{k}\Gamma(k,k')z_{k'}$,
which, in the case of generalized voter dynamics, 
Eq.~(\ref{eq:1}), becomes
\begin{equation}
  N_{\mathrm{eff}}=N
  \frac
  {
    \langle f(k)\rangle \langle k \rangle
    \left\langle\frac{kb(k)}{f(k)a(k)}\right\rangle^2 
  }
  {
    \left\langle\left\langle s(k,k') k
        b(k)\frac{[k'b(k')]^2}{f(k')a(k')}\right\rangle\right\rangle},
  \label{eq:2} 
\end{equation}
where we have defined
$\langle\langle\cdot\rangle\rangle=\sum_{kk'}P(k)P(k')(\cdot)$.

\section{Particular cases}
\label{sec:particular-cases}

With the formalism developed above, we can easily recover several of
the variations of the voter model proposed in the past. Let us look at
some of them in following section.

\subsection{Standard voter model and Moran process}
\label{sec:standard-voter-model}

The standard voter model and Moran process can be recovered by setting
$a(k)=b(k)=f(k)=s(k,k')=1$ and $a(k)=k$, $b(k)=k^{-1}$ and
$f(k)=s(k,k')=1$, respectively. In this case the known results
\cite{Sood08,PhysRevLett.94.178701} are recovered. Thus, for the voter
model, the conserved quantity is
\begin{equation}
  \label{eq:4}
  \omega= \sum_{k'}   \frac{k' P(k')}{\av{ k}} x_{k'}(t),
\end{equation}
the exit probability starting from a single $+1$ vertex is
\begin{equation}
  \label{eq:5}
  E_1(k) = \frac{k}{N \av{k }},
\end{equation}
and the consensus time 
takes the form
\begin{equation}
  \label{eq:6}
  T_N(\omega) = -N
  \frac{\av{k}^2}{\av{k^2 }} \left[ \omega
    \ln 
    \omega + (1-\omega) \ln (1-\omega) \right].
\end{equation}
On the other hand, for the Moran process we have
\begin{eqnarray}
  \label{eq:7}
  \omega &=&  \frac{1}{\av{k^{-1}}} 
  \sum_k \frac{P(k)}{k} x_k \\
  E_1(k)& =& \frac{1}{k} \frac{1}{N\av{k^{-1}}} \label{eq:14}\\
  T_N(\omega) 
 &=&  -N \av{k} \av{k^{-1}} \times \left[ \omega \ln(\omega) + (1-\omega) \ln(1-\omega)
   \right]. 
\end{eqnarray}
The difference between voter and Moran dynamics is quite evident here.
By looking the expressions for the conserved quantities,
Eq.~(\ref{eq:4}) states that for the voter model densities are
weighted with a factor $k/\langle k \rangle $ which compensates the
tendency of small degree nodes to change their state, whereas in the
Moran process the exact opposite occurs, being the density in
Eq.~(\ref{eq:7}) balanced by the factor $k^{-1}/\langle k^{-1}
\rangle$. This fact translates in the different forms of the exit
probability, Eqs.~(\ref{eq:5}) and~(\ref{eq:14}). For the voter model,
a single \textit{mutant} opinion can spread to the whole system more
easily when it first starts in a vertex of large degree, due to the
fact that large degree vertices are copied from with larger
probability. On the other hand, a \textit{mutant} in the Moran process
is able to spread faster if it starts on a low degree vertex, owing to
the corresponding fact that low degree vertices are invaded with low
probability.

\subsection{Voter model on weighted networks}
\label{sec:voter-model-weighted}

The extension of voter model to weighted networks
\cite{2010arXiv1011.2395B} is motivated by those situations in which the
strength of a relation can play a role in the process of opinion
formation. In this sense, weights would reflect the evidence that
the opinion of a given individual can be more easily influenced by a
close friend rather than by a casual acquaintance. In a weighted network, the voter model is
defined as follows: At each time step a vertex $i$ is
selected randomly with uniform probability; then one among the nearest neighbors
of $i$, namely $j$, is chosen with a probability proportional to the
weight $w_{ij} \ge 0$ of the edge joining $i$ and $j$. That is, the
probability of choosing the neighbor $j$ is
\begin{equation}
  P_{ij} = \frac{w_{ij}}{\sum_p w_{ip}}.
  \label{eq:8}
\end{equation}
Vertex $i$ is finally updated by copying the state of vertex $j$. If the 
weights depend on the degree of the edge's endpoints,
$w_{ij} = g(k_i, k_j)$, with $g(k,k')$ a symmetric multiplicative
function, i.e. $g(k,k') = g_s(k) g_s(k')$ \cite{2010arXiv1011.2395B},
voter dynamics is recovered by setting $s(k,k')=a(k) = f(k)=1$ and
$b(k) = g_s(k) \av{k}/\av{kg_s(k)}$, which leads to an invasion exit
probability 
\begin{equation}
  \label{eq:9}
  E_1(k) = \frac{k g_s(k)}{N \av{k g_s(k)}},
\end{equation}
and a consensus time
\begin{equation}
  \label{eq:10}
  T_N(\omega) = -N
  \frac{\av{k g_s(k) }^2}{\av{k^2 g_s(k)^2}} \left[ \omega
    \ln 
    \omega + (1-\omega) \ln (1-\omega) \right],
\end{equation}
with the conserved quantity
\begin{equation}
  \label{eq:11}
  \omega= \sum_{k'}   \frac{k' g_s(k')P(k')}{\av{ k
      g_s(k)}} x_{k'}(t) .
\end{equation}

In order to provide an example of weighted voter dynamics, we 
can consider the special case of weights scaling as a power law of the
degree, $g_s(k) = k^\theta$ on a scale-free network with degree
distribution of the form $P(k) \sim k^{-\gamma}$. The consensus time
starting from homogeneous initial conditions, $x_k(0) = 1/2$ takes the
form
\begin{equation}
  \label{eq:12}
  T_N(1/2) = N \ln(2) \frac{\av{k^{1+\theta} }^2}{\av{k^{2+2\theta}}}. 
  \label{FullT_N}
\end{equation}
From this expression, we can obtain different scalings with the
network size $N$, depending on the characteristic exponents $\gamma$
and $\theta$. Considering only $\gamma >2$ and using the scaling
behavior of the network upper cutoff $k_c \sim N^{1/2}$ for $\gamma<3$
and $k_c \sim N^{1/(\gamma-1)}$ for $\gamma>3$ \cite{mariancutofss},
we are led to different regions of behavior for $T_N(1/2)$:
%
%
\begin{enumerate}
\item If $\theta>\gamma-2$ both $\av{k^{1+\theta}}$ and
  $\av{k^{2+2\theta}}$ diverge. In particular, $\av{k^{1+\theta}} \sim
  k_c^{2+\theta-\gamma}$ and $\av{k^{2+2\theta}}\sim
  k_c^{3+2\theta-\gamma}$. Thus
  \begin{equation}
    T_N \sim  N k_c^{1-\gamma}.
  \end{equation}
  If $\gamma<3$, $k_c \sim N^{1/2}$, and $T_N \sim
  N^{(3-\gamma)/2}$. If $\gamma>3$, then $k_c \sim N^{1/(\gamma-1)}$,
  and $T_N \sim \mathrm{const}$.

\item If $\gamma-2 > \theta > (\gamma-3)/2$, then $\av{k^{1+\theta}}$
  converges and $\av{k^{2+2\theta}}$ diverges. Thus
  \begin{equation}
    T_N \sim  N  k_c^{\gamma-2\theta-3}.
  \end{equation}
  If $\gamma<3$, $T_N \sim N^{(\gamma-2\theta-1)/2}$; if $\gamma>3$,
  then $T_N \sim N^{2(\gamma-\theta-2)/(\gamma-1)}$.

\item If $\theta < (\gamma-3)/2$, then both $\av{k^{1+\theta}}$ and
  $\av{k^{2+2\theta}}$ converge, and we have
  \begin{equation}
    T_N \sim N.
  \end{equation}
\end{enumerate}


These scaling relations contain several interesting aspects. Among
these, it is worth highlighting that, in region $1$, and for
$\gamma>3$, the analytical results predict a constant scaling for the
consensus time, which therefore does \textit{not} depend on the
population size. As a consequence, in the thermodynamic limit the
ordering process is instantaneous. However, this turns out to be true
only at the mean field level, i.e. on annealed networks. Simulating
the process on quenched networks produces in fact different results,
with a clear dependence of the consensus time on $N$
\cite{2010arXiv1011.2395B}. Henceforth, this is a typical case showing
the limits of the mean field approach in predicting the behavior of
phenomena occurring on true finite networks with fixed connections and
disorder. Another interesting feature concerns the special value
$\gamma=3$. When $\theta > 0$, this value separates distinct scaling
behavior, while as $\theta<0$ it ceases to be a frontier, the scaling
of the consensus being linear in $N$ on both sides of it.

\section{A practical example: variable opinion strengths} 
\label{sec:pract-exampl-which}
The strength of the proposed generalized formalism resides in the
possibility of deriving simple HMF solutions to new variations of the
standard voter-like models, which were not studied in the past and whose
solution would have been too hard to compute with more exact techniques.
As an application of our formalism, we consider the case of voter
dynamics in a society in which certain individuals are more likely to
align their opinions with those of their neighbors than others
\cite{Antal06}. In particular, the propensity of a certain individual
to change opinion will depend on the strength of his/her social ties,
that is on his/her number of neighbors. We can accomplish such a
description in our formalism, by encoding this dependence in the
fitness function $f(k)$ and assuming $a(k)=b(k)=s(k,k')=1$ for the
sake of simplicity.  The HMF solution to such a problem will then be
easy to derive. Following the steps illustrated in the previous
sections, the conserved quantity reads
 \begin{equation}
 \omega=\sum_k P(k) \frac{k/f(k)}{\langle k/f(k)\rangle}x_k,
 \end{equation} 
 the exit probability for an individual seed in state $+1$
 \begin{equation}
 E_1= \frac{k/f(k)}{N\langle k/f(k)\rangle},
 \end{equation} 
 and the consensus time starting from a homogeneous state 
 $\omega$
 \begin{equation}
T_N(\omega) =-N  \frac{  \langle f(k) \rangle \langle k/f(k) \rangle^2 }
{ \langle k^2/   f(k) \rangle }
[\omega \ln\omega+(1-\omega)\ln(1-\omega)].
\end{equation}
 
\begin{figure}[t]
  \centering
    \includegraphics*[width=8cm]{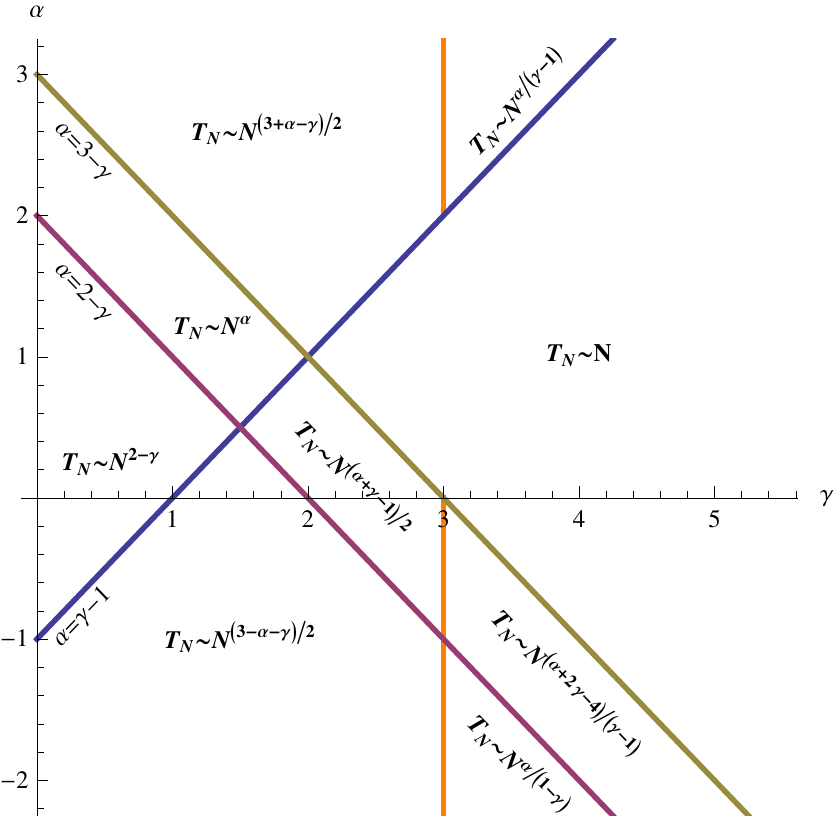}
    \caption{Phase diagram of the heterogeneous voter model with
      variable opinion strengths, in which the strength of the opinion
      of an agent (its willingness to change opinion) is related to
      its degree by the relation $f(k) = k^\alpha$.  }
  \label{fig:phasediagram2}
\end{figure}

We can provide a practical example by choosing $f(k)=k^\alpha$.  If
$\alpha<0$, less connected individuals are more likely to change
opinion, and connectedness can be interpreted as a measure of social
self-assurance. If $\alpha>0$, more connected individuals appear more
vulnerable to opinion variability. Interestingly, the standard voter
model is recovered for $\alpha=0$, where no agent heterogeneity is
postulated.  The conserved quantity assumes the simple form
$\omega=\langle k^{1-\alpha} x_k\rangle/ \langle k^{1-\alpha} \rangle$
and the consensus time starting from an initial condition $\omega=1/2$
is then
\begin{equation}\label{eq:TNhetero}
  T_N(1/2) =N\ln(2)   \frac{   \langle k^\alpha  \rangle  \langle k^{1-\alpha}
    \rangle^2  }{   \langle k^{2-\alpha} \rangle   }.
\end{equation}
The size scaling behavior of the consensus time can be derived from
Eq.  ~(\ref{eq:TNhetero}), following the procedure devised in the
previous section for the voter model on weighted networks. The results
are illustrated in compact form in
Fig. \ref{fig:phasediagram2}, where a phase diagram for the variables 
$\gamma$ and $\alpha$ is shown.  
The phase diagram conveys great insight into the role played by the 
parameter $\alpha$. Focusing on the region with $\gamma >2$, 
which is of greater interest in the study of dynamical processes on
complex networks, we obseve that for $3-\gamma < \alpha <\gamma -1$
the simple scaling relation $T_N\sim N$ is recovered. 
In analogy with a similar phenomenon observed
in the case of weighted networks in Section \ref{sec:voter-model-weighted},
the value $\gamma=3$ does not act as frontier and the same scaling
law is observed, regardless of the value of $\gamma$. Larger 
values of $\alpha$ break the balance that ensues the $T_N\sim N$ behavior
and lead to non-trivial size dependence of the consensus time, with 
exponents that depend on the degree distribution exponent $\gamma$.
If we restrict our analysis to the case of scale-free networks, corresponding 
to the $2<\gamma <3$ strip in the phase diagram,
we can easly see from the results in Figure \ref{fig:phasediagram2}
 that values of $\alpha$ in the range 
$| \alpha |<\gamma -1$  lead to either linear or sub-linear size scaling
of $T_N$, whereas for $| \alpha | > \gamma -1$ super-linear scaling is
encountered. This translates into the simple observation that in scale-free
networks, {\it large} degree selectivity makes consensus harder
to reach in larger systems, regardless of the sign of $\alpha$, i.e., 
of whether the degree selectivity makes low-degree individuals or high-degree
individuals more vulnerable to opinion change.

In order to corroborate our predictions, we have performed numerical
simulations of the voter dynamics at hand in complex networks with
power-law distributed degrees, generated with the Uncorrelated Model
(UCM) \cite{ucmmodel}.  We check the scaling behavior of the consensus
time, $T_N \sim N^{\beta}$, for three pairs of ($\gamma$,
$\alpha$) values of the phase diagram, located in the regions where 
super-linear, linear and sub-linear scaling of $T_N$ are observed
respectively.
 
\begin{figure}[t]
  \centering
  \includegraphics*[width=10cm]{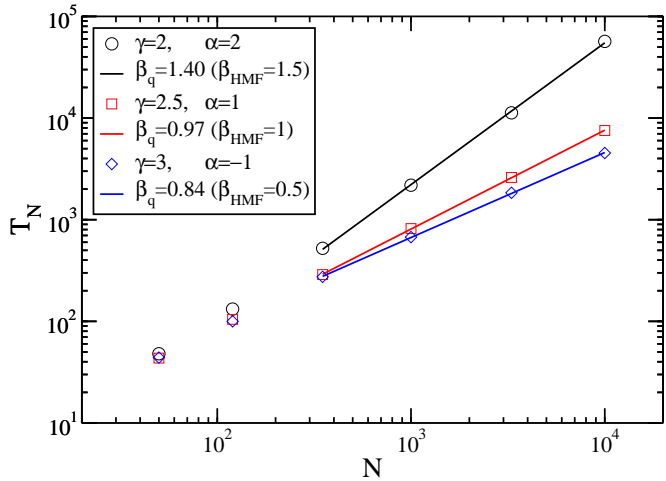}
   \caption{Scaling behavior of the consensus time $T_N$ as a function of the network size $N$ for different values of $\gamma$ and $\alpha$.     }
  \label{fig:T_N}
\end{figure}

In Fig. \ref{fig:T_N} we compare the scaling exponent predicted by
Eq. ~(\ref{eq:TNhetero}), $\beta_{\mathrm{HMF}}$, with the exponent $\beta_{q}$
obtained by fitting the numerical simulations run on quenched
networks.  The results are summarized in the same
Fig. \ref{fig:T_N}. We note that the HMF theory predicts
\textit{qualitatively} well the behavior of the voter dynamics, in the
sense that the scaling of $T_N$ with $N$ is super-linear for
$\gamma=2$ and $\alpha=2$, linear for $\gamma=2.5$ and $\alpha=1$, and
sub-linear for $\gamma=3$ and $\alpha=-1$, as expected. As for the
exact values of the $\beta$ exponents, the HMF prediction is
\textit{quantitatively} accurate only is certain regions of the phase
diagram, where the annealed-network approximation appears to
hold. Elsewhere, quenched-network effects take over and sensible
deviations with respect to the HMF value of $\beta$ are encountered in
simulations, in analogy with results for the weighted voter model, as
discussed in Section \ref{sec:voter-model-weighted}.

\section{Conclusions}
\label{sec:conclusions}
In this Chapter, we have presented a generalized model of consensus
formation, which is able to encompass all previous formulations of
copy/invasion processes inspired by variations on the voter model and
the Moran process. We considered the implementation of such
generalized dynamics on a heterogeneous contact pattern, represented
by a complex network, and derived the theoretical predictions for the
relevant dynamical quantities, within the assumptions of the
heterogeneous mean-field theory. We provided a brief review of
previous results that can be recovered by our generalized formalism,
and finally we considered a novel application to the case of opinion
formation in a social network. In particular, we addressed the case in
which the opinion strength of an individual is related to his/her
degree centrality in the network. We found that in scale-free networks
strong selectivity rules (which make less connected individuals much
proner to change their opinions than more-connected ones or vice
versa) lead to a steeper growth of consensus time with the system
size, making the ordering process slower in general. Numerical
simulations on quenched networks show that the HMF theory is able to
predict such behavior with reasonable accuracy. Slight deviations from
the theoretical predictions are encountered in certain regions of the
phase diagram, but they are due to quenched-network effects that the
HMF theory is not be able to capture.

\section*{Acknowledgements}

We acknowledge financial support from the Spanish MEC (FEDER) under
Projects No. FIS2010-21781-C02-01 and FIS2009-08451, and the Junta de
Andaluc\'{\i}a, under Project No. P09-FQM4682. R.P.-S. acknowledges
additional support through ICREA Academia, funded by the Generalitat
de Catalunya.

\bibliographystyle{spphys}
\bibliography{heterorev}

\printindex
\end{document}